\documentclass[Journal, 10pt]{IEEEtran}

\usepackage{color}
\usepackage{float}
\usepackage{tabularx,colortbl}
\usepackage{setspace}
\usepackage{cite,graphicx,amsmath,psfrag,bm}
\usepackage{subfigure}
\usepackage[usenames,dvipsnames]{xcolor}
\usepackage{mathabx}
\usepackage{cancel}
\usepackage{epstopdf}
\usepackage[process=auto]{pstool}
\usepackage{caption}
\usepackage[normalem]{ulem}
\graphicspath{{Figs/}}

\newcommand{\figref}{Fig.~\ref}
\newcommand{\tabref}{Table.~\ref}

\begin{document}

\title{A Simple Picosecond Pulse Generator Based on \\a Pair of Step Recovery Diodes}

\author{Lianfeng Zou,~ Shulabh Gupta,
        Christophe Caloz
\thanks{L. Zou (lianfeng.zou@polymtl.ca) and C. Caloz are with Polytechnique Montr\'{e}al, Montr\'{e}al, Qu\'{e}bec, Canada.

S. Gupta is with Carleton University, Ottawa, Ontario, Canada.}%
%\thanks{Manuscript received April 19, 2005; revised December 27, 2012.}
}
%\markboth{IEEE Microwave and Wireless Components Letters ,~Vol.~, No.~, Month~2016}%
%{Shell \MakeLowercase{\textit{et al.}}: Bare Demo of IEEEtran.cls for Journals}

\maketitle

\begin{abstract}
A picosecond pulse generator based on a pair of step recovery diodes (SRD), leveraging the transient response of the SRD PN junction and controlling the pulse width by a resistor, is proposed. We first explain the operation principle of the device, decomposing the pulse generation into different phases, and then demonstrate an experimental prototype with two different resistance, and hence pulse width, values.
\end{abstract}

\begin{IEEEkeywords}
UWB pulse generator, real-time analog signal processing (R-ASP), step recovery diode (SRD)
\end{IEEEkeywords}

\IEEEpeerreviewmaketitle
%\tableofcontents
\section{Introduction}

The emerging Real-time Analog Signal Processing (R-ASP) technology~\cite{Jour:2013_MwMag_Caloz}, using dispersion-engineered structures (analog signal processors) providing specified group delays versus frequency responses for processing high-frequency and wide-bandwidth signals in real time, may lead to novel low-cost and low-complexity radio systems~\cite{CONF:2016_APS_ZOU} for future millimeter wave and terahertz wireless communications.

Analog signal processors usually deal with ultra-short pulses with very rich spectral contents. Such pulses may be generated by different techniques~\cite{REPORT:2008_PPL}. Among them, the step recovery diode (SRD) approach~\cite{JOUR:2014_MOTL_Kamal,JOUR:2001_TMTT_Lee,JOUR:20012_TIM_Xia} features simple circuit design, while providing picosecond pulse generation capability. Conventional SRD-based pulse generators usually employ an SRD to generate a pulse with picosecond rising edge, which destructively interferes with its opposite-polarity delayed replica produced by a short-circuited stub, which results in a short pulse whose width is equal to the round trip time along the stub. However, the stub configuration suffers from spurious reflections in both forward and backward directions, whose suppression requires pulse shaping networks~\cite{JOUR:20012_TIM_Xia}, and requires complicated tuning schemes~\cite{CONF:2006_IMS_Zhang}.

In this paper, we propose a stub-less pulse generator based on a pair of SRDs, which is immune of spurious reflections and which provides flexible control of the pulse width via a simple resistance.

\section{Principle}
\begin{figure}[h!t]
   \centering
   \includegraphics[width=1\columnwidth]{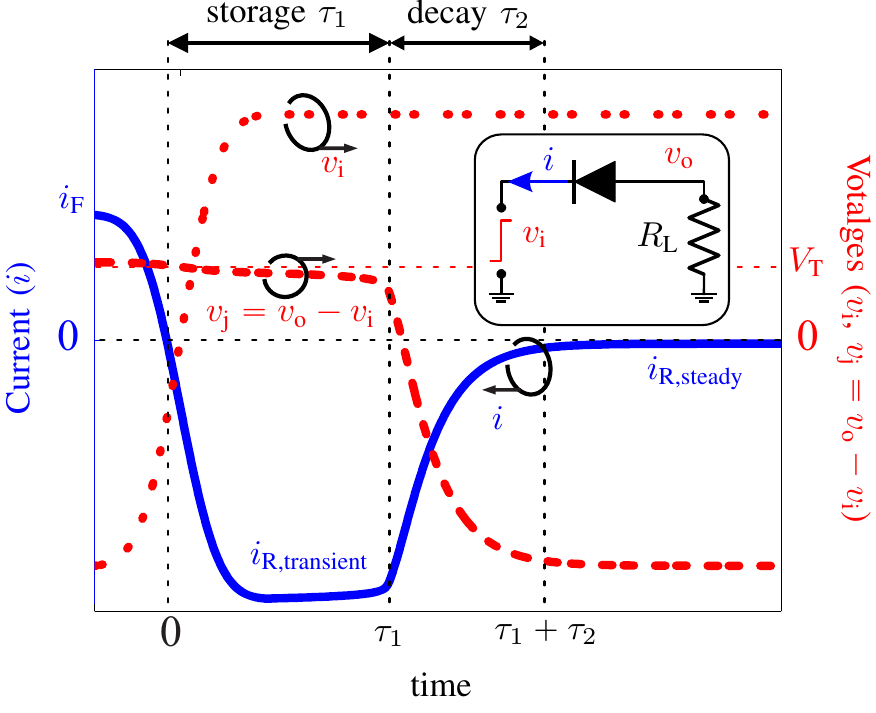}
   \caption{PN junction transient current $i$ (solid blue) and junction voltage $v_\text{j}$ (dash red) in response to a step voltage $v_i$ (dot red) from the forward to the reverse bias regimes.}
   \label{FIG:Transient}
\end{figure}

The transient behaviour of a PN junction in response to a step voltage from the forward to the reverse bias regimes is shown in~\figref{FIG:Transient}. Initially, the PN junction is forward biased with current $i_\text{F}$. When the voltage $v_\text{i}$ reverses polarity, the junction is not switched off immediately, but rather takes time $\tau_1$ (storage phase), with current $i_\text{R,transient}$, to neutralize the charges stored in the depletion region and progressively raise the junction barrier~\cite{BK:2005_Tse}. Note that during the whole storage phase, the junction maintains low resistance, and almost constant junction voltage, i.e. $v_\text{j} \approx V_\text{T}$. Near the end of the storage phase, the raising junction barrier increases the resistance, finally switching off the junction for a decay time $\tau_2$ (decay phase), which is extremely short (in picosecond range) for an SRD~\cite{BK:2005_Tse}. Assuming constant $i_\text{F}$ and constant $i_\text{R,transient}$, the storage phase time is given by~\cite{BK:2005_Tse}
\begin{equation}\label{EQ:TAU1}
\tau_1 = \tau_{L}\ln{\left(1+\dfrac{i_\text{F}}{i_\text{R,tran}}\right)},
\end{equation}
where $\tau_{L}$ is the minority carrier lifetime. We shall next present our pulse generator based on this SRD transient physics.

\begin{figure}[h!t]
   \centering
   \includegraphics[width=0.9\columnwidth]{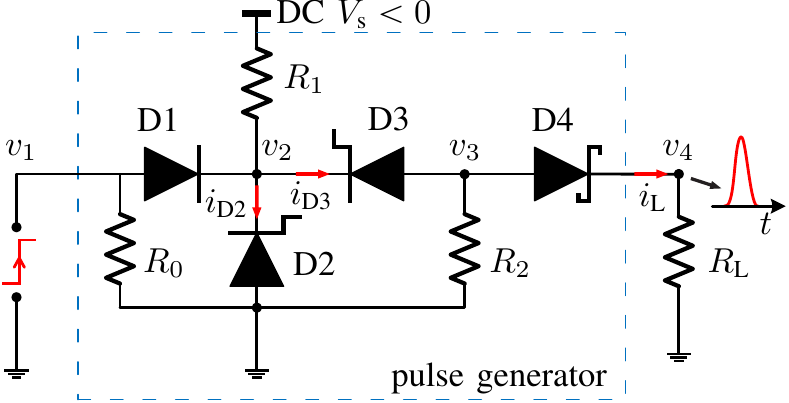}
   \caption{Proposed SRD-pair pulse generator.}
   \label{FIG:Schematic}
\end{figure}

\begin{figure}[h!t]
\centering
\subfigure[]{
\includegraphics[width=0.78\columnwidth]{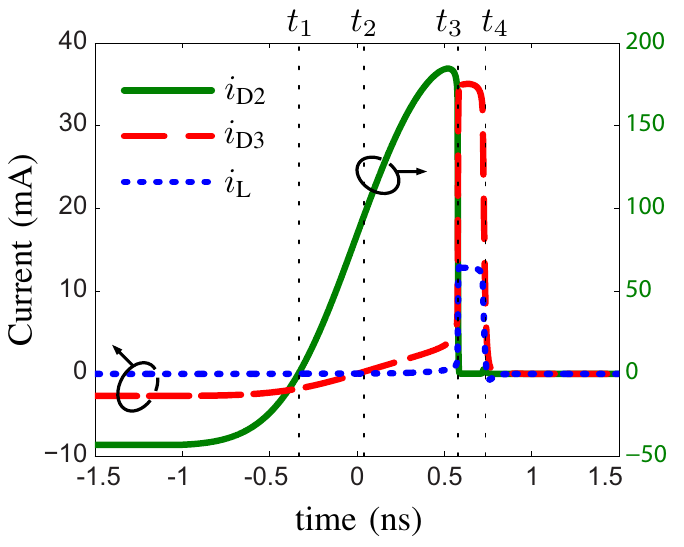}
}
\subfigure[]{
\includegraphics[width=0.78\columnwidth]{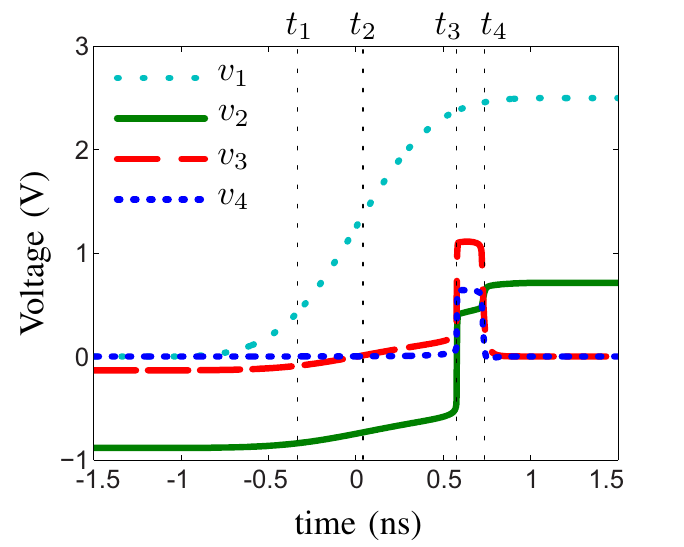}
}
   \caption{Circuit simulated (a) currents (polarities referred to current directions indicated in~\figref{FIG:Schematic}) and (b) voltages, with \mbox{$V_\text{s}=-3.5$~V}, $R_1=25$~$\Omega$ , $R_2=50$~$\Omega$, and Spice models for the diodes in~\tabref{TAB}.  }
   \label{FIG:Waveforms}
\end{figure}

Figure~\ref{FIG:Schematic} shows the schematic of the proposed SRD-pair pulse generator. The generator is triggered by the rising edge of the input voltage through PIN diode D1, that is always ON for blocking possible negative input voltages. Its core is the parallel connection of SRDs D2 and D3, biased at their cathode by DC $V_s<0$ through $R_1$ current limiter, with $R_2$ controlling the pulse width, as will be shown later. Finally, Schottky diode D4, operates as a pulse formation switch. The resistor $R_0=50$~$\Omega$ ensures input matching during the pulse production. Assuming square-wave (0-2.5~V) excitation $v_1$, the operation may be decomposed into 5 phases, shown in~\figref{FIG:Waveforms}.

\begin{table}[h!]
\begin{center}
\caption{Diodes used in the generator with reference to~\figref{FIG:Schematic}.\\}
\label{TAB}
  {\begin{tabular}{|c|c|}\hline

Notation & Model                    \\\hline
D1      & Avago Tech. HMPS-2822     \\\hline
D2      & Aeroflex MMDB30-0805      \\\hline
D3      & Aeroflex MMDB30-0805      \\\hline
D4      & Skyworks SMS7630-061-0201 \\\hline
\end{tabular}}
\end{center}
\end{table}

$\bullet$ Phase 1 ($t<t_1$): D2 and D3 are forward biased, and D4 is off. The excitation $v_1$ is close to 0~V, so the D2 and D3 forward currents are such that $i_\text{F,D2}\gg i_\text{F,D3}$ due to the existence of $R_2$ in the D3 loop. The junction voltages $v_\text{j,D2} =-v_2$ and $v_\text{j,D3} =v_3-v_2$, where $v_3<0$ due to the current flowing in $R_2$ from the ground to node 3, so that D4 is off and $v_4=0$.

$\bullet$ Phase 2 ($t_1<t<t_2$): D2 is in storage regime, D3 is still forward biased, and D4 is still off. The excitation $v_1$ starts rising, driving D2 ($i_\text{R,D2}>0$) into storage regime before D3 due to the smaller resistance in the D2 current loop than that in the D3 one and hence faster current changing rate ($\partial i_\text{D2}/\partial t>\partial i_\text{D3}/\partial t$) in response to the rising edge of $v_1$. According to~\figref{FIG:Transient}, $v_\text{j,D2}=-v_2$ varies only slightly in storage regime, and $v_3<0$ because D3 is still forward biased, and hence keeping D4 off.

$\bullet$ Phase 3 ($t_2<t<t_3$): D2 and D3 are both in storage regime, D4 is still off. The voltage $v_1$ eventually also drives D3 into storage regime ($i_\text{R,D3}>0$). Since $i_\text{F,D2}/i_\text{R,D2}\ll i_\text{F,D3}/i_\text{R,D3}$, we have $\tau_{1,\text{D2}}<\tau_{1,\text{D3}}$ according to~\eqref{EQ:TAU1}, and therefore D2 will switch off before D3. In the meanwhile, $v_3$ follows $v_2$, with insufficient variation to switch D4 on.

$\bullet$ Phase 4 ($t_3<t<t_4$): D2 is off, D3 is still in storage regime, D4 is on. The SRD D2 abruptly switches off at $t_3$, while D3 remains in storage regime. Consequently, $v_2$, and following $v_3$, abruptly increase, which switches D4 on, and therefore generates at the output the rising edge of $v_4$. Moreover, the off-switching of D2, and the on-switching of D4, which makes $R_\text{L}$ parallel to $R_2$, together boost $i_\text{R,D3}$, starting to shorten the D3 storage time, which would otherwise be longer as expected in Phase 3.

$\bullet$ Phase 5 ($t>t_4$): D2, D3, D4 are all off. The D3 storage phase suddenly ends at $t_4$, nullifying $i_\text{R,D3}$ and $v_3$, and hence switching D4 off, which results in the falling edge of $v_4$.

When the input level goes back to zero, the generator returns to the regime of Phase~1, waiting for the next voltage step to resume the aforementioned pulse formation cycle.

\begin{figure}[h!t]
   \centering
      \includegraphics[width=1\columnwidth]{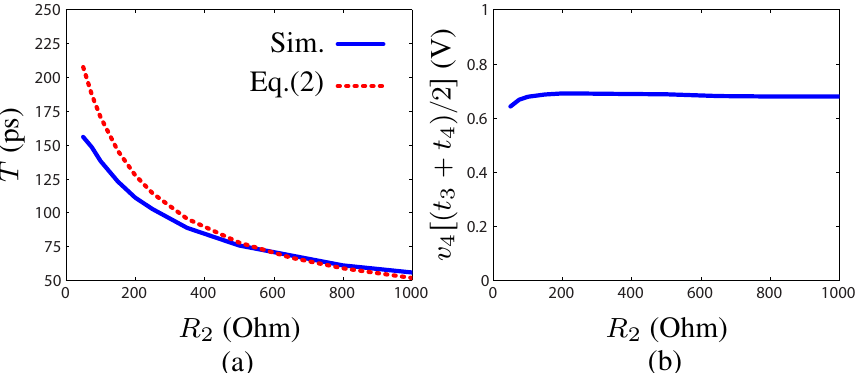}
   \caption{(a) Circuit simulated half-magnitude pulse width (solid blue) and calculated pulse width (dash red) by \eqref{EQ:PLSW1} (with simulated currents and $\tau_\text{L}=3$~ns~\cite{DS:2005_SRD}) versus $R_2$. (b)~Circuit simulated pulse magnitude. }
   \label{FIG:WMVR}
\end{figure}

Let us now examine the factors determining the pulse width. Equation~\eqref{EQ:TAU1} assumes constant $i_\text{R,transient}$ over the entire storage phase $\tau_1$. As explained above, the storage phase for D3 is from $t_2$ to $t_4$. However, since $i_\text{R,D3}(t_2<t<t_3)\ll i_\text{R,D3}(t_3<t<t_4)$, the storage time essentially reduces to the interval $[t_3,t_4]$, and $i_\text{R,transient}$ in~\eqref{EQ:TAU1} should be replaced by $i_\text{R,D3}(t_3<t<t_4)$ for the effective storage time of D3. Since this interval ($[t_3,t_4]$) corresponds to the pulse duration, this duration can be approximated as
\begin{equation}\label{EQ:PLSW1}
  T  \approx \tau_{L}\ln{\left[1+\dfrac{i_\text{F,D3}}{i_\text{R,D3}(t_3<t<t_4)}\right]}.
\end{equation}
with an accuracy that is proportional to $i_\text{R,D3}(t_2<t<t_3)$ and hence inversely proportional to $R_2$, i.e. the closer $i_\text{R,D3}$ to zero in the interval $[t_2, t_3]$ (associated with larger $R_2$), the better the accuracy, as will be shown later in~\figref{FIG:WMVR}(a).
%For instance, reading out from~\figref{FIG:Waveforms}, \mbox{$i_\text{F,D3}=2.5$~mA}, $i_\text{R,D3}(t_3<t<t_4)=37$~mA, with $\tau_{L}=3$~ns~\cite{DS}, Eq.~\eqref{EQ:PLSW1} yields \mbox{$T=196$~ps}, which is just a bit larger than the half-magnitude width (165~ps) in~\figref{FIG:Waveforms}.
Using Ohm law, one may further write~\eqref{EQ:PLSW1} in the form
\begin{equation}\label{EQ:PLSW2}
  T   \approx \tau_{L}\ln{\left(1+r\dfrac{R_2||R_\text{L}}{R_2}\right)} =  \tau_{L}\ln{\left(1+r\dfrac{R_\text{L}}{R_2+R_\text{L}}\right)},
\end{equation}
where $r$ is the ratio of $v_{3}$ in the forward bias regime to $v_{3}$ in the $[t_3,t_4]$ interval. Equation~\eqref{EQ:PLSW2} reveals that, for given $R_\text{L}$, one may adjust $R_2$ to control the pulse width. This is confirmed by the simulation result in~\figref{FIG:WMVR}(a), which also show the result for the approximation in~\eqref{EQ:PLSW1}. Finally, Fig.~\ref{FIG:WMVR}(b) plots the pulse magnitude, which is seen to be essentially flat versus $R_2$.

\section{Experimental Demonstration}

\begin{figure}[h!t]
   \centering
   \includegraphics[width=0.8\columnwidth]{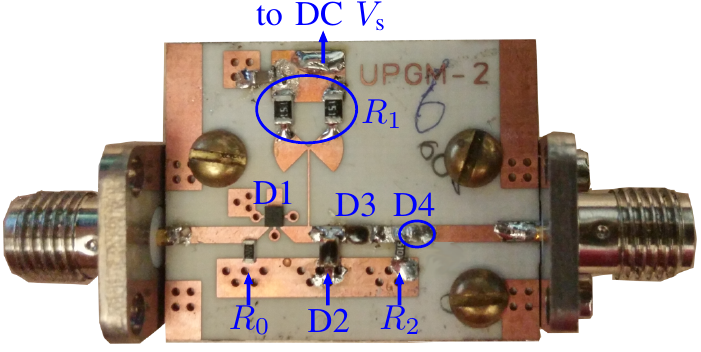}
   \caption{Experimental prototype.}
   \label{FIG:Hardware}
\end{figure}

\begin{figure}[h!t]
   \centering
   \subfigure[]{
   \includegraphics[width=0.75\columnwidth]{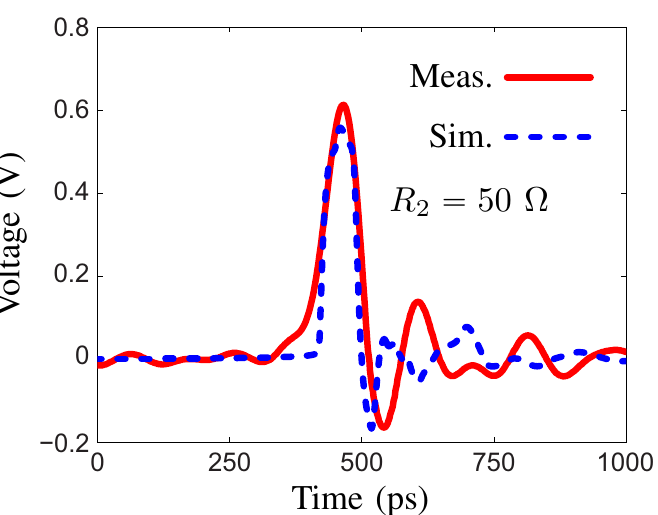}
    }
   \subfigure[]{
   \includegraphics[width=0.75\columnwidth]{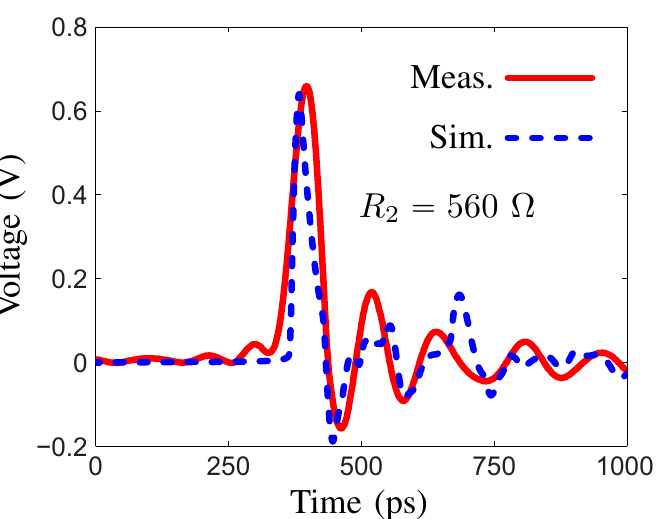}
   }
   \caption{Experimental (solid red) and circuit simulation (with finite-length transmission lines) (dash blue) results corresponding to (a)~$R_2=50$~$\Omega$ and (b)~$R_2=560$~$\Omega$ , ($V_\text{s}=-3.5$~V and $v_\text{i}=0/2.5$~V).}
   \label{FIG:MEAS}
\end{figure}

Figure~\ref{FIG:Hardware} shows the fabricated prototype, whose diode models are listed in~\tabref{TAB}, and where $R_1$ consists in two 0805 150~$\Omega$ resistors in parallel, $R_0$ is a 0603 50~$\Omega$ resistor, and $R_2$ is a 0603 50 or 560~$\Omega$ resistor. The experimental results are shown in~\figref{FIG:MEAS} (solid red). The half-magnitude pulse widths are $94$ and $62$~ps for $R_2=50$~$\Omega$ and $560$~$\Omega$, respectively, which are different from the corresponding simulation pulse widths in~\figref{FIG:WMVR}(a). Ringing tails are also observed. To understand the cause of the pulse width discrepancy between simulation and experiment and the experimental ringing tails, one should note that the transmission line lengths are assumed to be zero in~\figref{FIG:Schematic} and in the corresponding simulation results in~\figref{FIG:Waveforms} and~\figref{FIG:WMVR}. However, in the fabricated prototype, nonzero-length transmission lines delay the multiple internal reflections between diodes, which are not perfectly matched, hence generating successive replicas of the intended output pulse corresponding to the ringing tail. Moreover, the spurious pulses may constructively or destructively interfere with the main one, resulting in wider or shorter pulse width compared to the simulated on in~\figref{FIG:WMVR}(a). One may confirm this explanation by including finite-length transmission lines in the simulation circuit, as shown by the dash-blue curves in~\figref{FIG:MEAS}. In practice, one may naturally improve the output pulse shape by enhancing the matching of the diodes and using the shortest possible interconnecting lines.

\section{Conclusion}

A simple pulse generator based on a pair of SRDs with pulse width adjustable by a single resistor is presented, analyzed and experimentally demonstrated. Such a pulse generator may find wide applications in R-ASP and other UWB radio systems with low cost and low complexity.

\bibliographystyle{IEEEtran}
%\nocite{*}
\bibliography{IEEEabrv,ref}

\end{document}